\documentclass[aps,prl,twocolumn]{revtex4}
\usepackage{amsfonts,amssymb,amsmath,bm}
\usepackage{times,color,graphicx}
\usepackage{subfigure}

\newcommand{\beq}{\begin{equation}}
\newcommand{\eeq}{\end{equation}}

\begin{document}

\title{Stochastic Optimal Control as Non-equilibrium Statistical Mechanics:\\
Calculus of Variations over Density and Current}
\author {Vladimir Y. Chernyak$^{a,b}$}
\author{Michael Chertkov$^{b}$}
\author{Joris Bierkens $^{c}$}
\author{Hilbert J. Kappen $^{c}$}

\affiliation{$^a$Department of Chemistry, Wayne State University, 5101 Cass Ave,Detroit, MI 48202, USA}

\affiliation{$^b$Center for Nonlinear Studies and Theoretical Division, LANL, Los Alamos, NM 87545, USA}

\affiliation{$^c$ SNN Adaptive Intelligence, Radboud University, Nijmegen
P.O. Box 9101, 6500 HB, Nijmegen, the Netherlands}

\date{\today}

\begin{abstract}
In Stochastic Optimal Control (SOC) one minimizes the average cost-to-go, that consists of the cost-of-control (amount of efforts), cost-of-space (where one wants the system to be) and the target cost (where one wants the system to arrive), for a system participating in forced and controlled Langevin dynamics. We extend the SOC problem by introducing an additional cost-of-dynamics, characterized by a vector potential. We propose derivation of the generalized gauge-invariant Hamilton-Jacobi-Bellman equation as a variation over density and current, suggest hydrodynamic interpretation and discuss examples, e.g., ergodic control of a particle-within-a-circle, illustrating non-equilibrium  space-time complexity.
\end{abstract}


\maketitle

In its standard setting the problem of the Stochastic Optimal Control (SOC) involves minimizing the average over stochastic trajectories of the cost-to-go, which consists of the cost of using the control field, the cost of arriving to a certain position $\bm{x}$ in the configuration space ${\cal M}$, described by a potential function $\varphi(\bm{x})$, and the cost accumulated along the trajectory, described by a time-dependent potential $V(\bm{x},\tau)$. The potentials $\varphi(\bm{x})$ and $V(\bm{x},\tau)$ can be viewed as variables dual to the particle probability distributions at the arrival time and during the time evolution, respectively, and thus interpreted as Lagrange multipliers.

In this Letter, continuing the thread of \cite{77Fle,81Mit,05Kap,07Tod,08Tod,11Kap} where methods of statistical and quantum mechanics were applied to SOC, we extend the standard cost-to-go functional by adding a term, associated with a vector potential, $\bm{A}(\bm{x},\tau)$, which leads to (i) a variational derivation of the stochastic Hamilton-Jacobi-Bellman (HJB) equation and its more general Gauge Invariant (GI) version, (ii) extend the capability of the HJB equation to treat such observables as produced work, generated entropy, and fluxes that occur in systems with non-contractible cycles in phase space.

Stochastic dynamics of a particle in a compact $m$-dimensional space ${\cal M}$, e.g., an $m$-dimensional torus, is described by the following Langevin equation
\begin{eqnarray}
\label{Langevin} \dot{\eta}^{j}\doteq\frac{d}{d\tau}\eta^j=f^{j}(\tau,{\bm \eta})+u^{j}(\tau,{\bm \eta})+\xi^{j}(\tau,{\bm \eta}),\\
\label{noise}\langle\xi^{j}(\tau,{\bm \eta})\rangle=0,\quad \langle\xi^{j}(\tau,{\bm \eta}) \xi^{k}(\tau',{\bm \eta})\rangle =\kappa g^{jk}\delta(\tau-\tau'),
\end{eqnarray}
where $\tau\in [t;T]$; ${\bm f}$ is the ``force" field, deterministic and assumed known; ${\bm u}$ is the ``control" field which as we will see below is subject to our optimization/choice; and ${\bm \xi}$ is a $\delta$-correlated in time, zero-mean Gaussian random field, whose correlations are fully expressed via a strictly positive symmetric matrix, $\kappa g({\bm \eta})$, where $\kappa$ measures the noise strength, and $g$ can be viewed as a (generally space-time dependent) metric in the configuration space ${\cal M}$, with $g_{ij}g^{jk}=g^{kj}g_{ji}=\delta_i^k$, where we use standard in theoretical physics covariant notations, i.e., assuming summation over repeating pairs of sub/superscripts, and applying the metric to relate vectors to co-vectors, e.g. $f_i=g_{ij}f^j$.

We consider a problem describing the optimal choice of the control vector field, ${\bm u}$, in Eq.~(\ref{Langevin})
\begin{widetext}
\begin{eqnarray}
&& {\cal C}(t,{\bm x};T)\doteq\min_{\{\bm u\}} C({\{\bm u\}};t,{\bm x};T),\quad C({\{\bm u\}};t,{\bm x};T)\equiv \Biggl\langle\varphi(\eta(T))+\int_t^T d\tau \left(\frac{1}{2} h_{ij} u^i u^j+V+A_j\dot{\eta}^j\right) \Biggr\rangle,\label{C}\\
&& \langle B(t',\eta(t'))\rangle\doteq\frac{\int_{{\bm \eta}(t)={\bm x}} {\cal D}{\bm \eta}
\exp\left(-\kappa\int_t^{t'}d\tau(\dot{\eta}^i-f^i-u^i)g_{ij}(\dot{\eta}^j-f^j-u^j)\right)
B(t',\eta(t'))}{\int_{{\bm \eta}(t)={\bm x}} {\cal D}{\bm \eta}
\exp\left(-\kappa\int_t^{t'}d\tau(\dot{\eta}^i-f^i-u^i)g_{ij}(\dot{\eta}^j-f^j-u^j)\right)}, \label{average}
\end{eqnarray}
\end{widetext}
where Eq.~(\ref{average}) defines averaging over stochastic trajectories in terms of a path integral; all co-vector and tensor fields in the integrand of Eq.~(\ref{C}) may depend explicitly on $\tau$ and $\bm{x}$; minimization/variation over ${\{\bm u\}}$ is functional,  i.e. we minimize over all $u(\tau,\eta(\tau))$;  and Eq.~(\ref{average}), stated as a path integral over ${\bm \eta}(\tau)$ defines averaging over the stochastic trajectories evolving according to Eq.~(\ref{Langevin}).

The meaning of the four terms under the average in the cost-to-go $C({\{\bm u\}};t,{\bm x};T)$ is as follows: the first local term describes  the target cost, i.e. the cost for the system to arrive at the final moment of time $T$ at ${\bm \eta}(T)$; the second term (which is also the first integral term) defines the cost-of-control; the third term stands for the cost-of-space, as it measures the cost depending on where the system stays in the phase space during the entire interval; finally,  the last term in $C({\{\bm u\}};t,{\bm x};T)$, as shown in Eq.~(\ref{C}), represents the cost-of-dynamics,  i.e. it is sensitive to how the system is moving in phase space during the period of interest (the cost is zero if the system does not move). The first three terms in Eq.~(\ref{C}) are standard in control theory, while the fourth term is new. It is also natural (exploiting theoretical physics jargon and intuition) to refer to $V$ and ${\bm A}$ in Eq.~(\ref{C}) as the scalar and vector potentials, respectively.
Obviously, the average cost-to-go ${\cal C}$ depends functionally on $V$ and ${\bm A}$.

Following \cite{09CCMT} we define the so-called average density and average current-density (hereafter referred to as just density and current) for the Langevin dynamics given by Eq.~(\ref{Langevin}):
\begin{eqnarray}
&& \rho(\{{\bm u}\};t,{\bm x};T)\doteq\int_t^T \frac{d\tau}{T-t}\langle\delta({\bm \eta}(\tau)-{\bm x})\rangle,
\label{rho}\\
&& {\bm J}(\{{\bm u}\};t,{\bm x};T)\doteq\int_t^T \frac{d\tau}{T-t} \langle\dot{\bm \eta}\delta({\bm \eta}(\tau)-{\bm x})\rangle.
\label{J}
\end{eqnarray}
In what follows we will simplify the notations, dropping the dependence of the density and current on $T$. Utilizing Eq.~(\ref{average}), and defining averaging over stochastic trajectories, one finds that the density satisfies the Fokker-Planck equation, which can also be interpreted as the continuity equation relating the current and density,
\begin{eqnarray}
\label{FP} \partial_t\rho +\sqrt{g}\partial_i\frac{g^{ij}}{\sqrt{g}}J_j=0,\quad J_i=-\kappa\partial_i\rho+(f_i+u_i)\rho,
\end{eqnarray}
where $g\doteq\det(g^{ij})$, and $\partial_i\doteq\partial_{x^i}$. Naturally, $\rho$ is nonnegative and, according to Eq.~(\ref{FP}), properly normalized at any time, $\int_{\cal M} d{\bm x} \rho/\sqrt{g}=1$. Note that the second equation in (\ref{FP}) can be inverted, thus expressing the control field ${\bm u}$ explicitly in terms of the current and density. Therefore, applying Eqs.~(\ref{rho},\ref{J},\ref{FP}) to Eq.~(\ref{C}),  one arrives at the following expression in terms of the density and current for the average (still not yet optimized with respect to ${\bm u}$) cost-to-go
\begin{eqnarray}
&& \hspace{-0.8cm} C\!=\!C_0\!+\!\int\limits_{\cal M}\!\frac{d{\bm x}}{\sqrt{g}}\Biggl(\varphi({\bm x})\rho(T,{\bm x})\!+
\!\int\limits_t^T \left(V\rho+g^{ij}A_iJ_j\right)\Biggr), \label{C-rho-J}\\
&& \hspace{-0.8cm} C_0\!\doteq\!\! \int\limits_{\cal M}\!\!\frac{d{\bm x}}{\sqrt{g}}
\!\int\limits_t^T\!\!d\tau
\frac{h^{ij}(\kappa\partial_i\rho\!+\!J_i\!-\!f_i\rho)(\kappa\partial_i\rho\!+ \!J_i\!-\!f_i\rho)}{2\rho},\label{C0}
\end{eqnarray}
where the average cost-of-control $C_0$ does not depend on the scalar and vector potentials, and is defined as an explicit functional of $\rho$, ${\bm J}$ and function of $t,{\bm x}$ and $T$, i.e. $C_0(\{\rho\},\{{\bm J}\};t,{\bm x};T)$. Thus, the optimization task in Eq.~(\ref{C}) translates into minimizing the r.h.s. of Eq.~(\ref{C-rho-J}) under the conditions of Eqs.~(\ref{FP}). Introducing a functional Lagrangian multiplier $\Phi(\tau,{\bf x})$  for the first equation in (\ref{FP}), adding the corresponding term to Eq.~(\ref{C-rho-J}), we compute the variations over ${\bm J}(\tau,{\bm x})$ and $\rho(\tau,{\bm x})$, under condition $\delta\rho(t;{\bm x})=0$ (the initial density is fixed). Combining the two variation equations results in the following closed form equation
\begin{eqnarray}
 \partial_\tau\Phi&=& \kappa\sqrt{g}\partial_{i}\frac{g^{ij}}{\sqrt{g}}(\partial_{j}\Phi+A_{j})+ g^{ij}f_{i}(\partial_{j}\Phi+A_{j}) \nonumber
\\ &+& V -\frac{1}{2} g^{ik}h_{kl}g^{lj}(\partial_{i}\Phi+A_{i})(\partial_{j}\Phi+A_{j})), \label{HJB}
\end{eqnarray}
which should be solved backwards in time with the ``initial'' condition $\Phi(T,{\bm x})=\varphi({\bm x})$ that originates from the contact term in the variation over $\rho(T)$.

A number of remarks with regards to Eq.~(\ref{HJB}) is in order. First, combining the equation emerging in the result of variation over $\bm{J}$ with the second relation in Eq.~(\ref{FP}) one arrives at the following explicit expression of the optimal control field $\bm{u}$ via the optimal $\Phi$
\begin{eqnarray}
u_i=-\bar{h}_i^k(\partial_k\Phi+A_k),
\quad \bar{h}_{ij}h^{jk}=\delta_i^k,\quad\bar{h}_i^k=\bar{h}_{ij}g^{jk}.
\label{u_opt}
\end{eqnarray}
Second, Eq.~(\ref{HJB}) is gauge invariant under simultaneous transformation of the scalar and vector potentials: $A_i\to A_i +\partial_i\phi, V\to V+\partial_\tau\phi$, where $\phi(\tau,{\bm x})$ is an arbitrary scalar function. Third, the Lagrangian multiplier solving Eq.~(\ref{HJB}) actually coincides with the optimal average cost-to-go function, ${\cal C}(t,{\bm x};T)=\Phi(t,{\bm x})$. The relation follows from multiplying Eq.~(\ref{HJB}) by $\rho$, integrating the result over the $d{\bm x}/\sqrt{g}$ and also over $\tau$ in the $[t;T]$ interval,  and then comparing the final expression with Eq.~(\ref{C-rho-J}). Fourth, the terms in Eq.~(\ref{HJB}) that contain the gauge field can be obviously absorbed into the force field $\bm{f}$ and scalar potential $V$, resulting in the celebrated stochastic -Hamilton-Jacobi-Bellman (HJB) equation of the control theory, which means that Eq.~(\ref{HJB}) is a particular case of the standard HJB equation, and access to an efficient solver of the latter provides a way to
solve Eq.~(\ref{HJB}). On the other hand, the equivalence between ${\cal C}$ and $\Phi$ means that one can also replace $\Phi$ in Eq.~(\ref{HJB}), thus discovering that Eq.~(\ref{HJB}) can be viewed as a Gauge Invariant (GI) generalization of the HJB, rather than a particular case, since it allows to consider control over a broader set of phenomena (e.g., work/entropy generation, fluxes, etc.). Finally, combining Eqs.~(\ref{HJB},\ref{u_opt}) we arrive at the following version of the GI-HJB equation (\ref{HJB}) stated in terms of the control field
\begin{eqnarray}
&-&\!\partial_\tau u_i
\!-\!\underbrace{\frac{1}{2}\bar{h}_i^k\partial_k h_{kl}h^{lm}h^{kn} u_m u_n}_{\mbox{"self-advection"}}
\!-\!\!\underbrace{\bar{h}_i^k\partial_k f_j h^{jm}u_m}_{\mbox{"advection by force"}}\nonumber\\ &=&\!
\underbrace{\kappa\bar{h}_i^k\partial_k\sqrt{g}\partial_lh^{ln}u_n/\sqrt{g}}_{\mbox{"dissipation"}}
\!+\!\!\underbrace{\bar{h}_i^k\left(\partial_\tau A_k-\partial_k V\right)}_{\mbox{"pumping/constraints"}}\!\!\!,
\label{HJB-u}
\end{eqnarray}
where, to avoid tedious expressions, we assumed that both $h$ and $g$ metrics are time-independent.

All terms in  Eq.~(\ref{HJB-u}) allow for a very natural hydrodynamic interpretation, where the optimal control field, ${\bm u}$, is interpreted up to a metric-dependent re-normalization as the ``velocity" field that evolves backwards in time in a compact space with curvature $g$. The second term on the l.h.s. of Eq.~(\ref{HJB-u}), also the only nonlinear term in the equation, describes the ``self-advection of velocity by itself". Continuing the hydrodynamic analogy, one interprets the second term on the l.h.s as ``advection by an external field ${\bm f}$". Then, the first term on the r.h.s. stands for the dissipation/viscosity, induced by the noise in the original Langevin equation (\ref{Langevin}). Finally, the last term on the r.h.s. represents pumping/injection, it may also represent constraints,  e.g. expressing relations between pressure and density, the phase space hydrodynamics. Details and consequences of the ultimate relation between the control and hydrodynamics will be discussed elsewhere \cite{
12CC}.

Scalar and vector potentials,  as well as $\varphi({\bm x})$,  can also be viewed as functional Lagrangian multipliers used to fix specific forms of the density and current functions. (Note, however, that in this formulation $\rho$ and $J$ are not fully arbitrary but consistent with each other through the continuity equation,  i.e. the first equation of Eqs.~(\ref{FP}).) Therefore, under fixed and consistent scalar and vector potentials no additional optimization in Eq.~(\ref{C}) is needed, the control field is completely defined by the second equation of (\ref{FP}),  and then the average cost is just the cost-of-control, $C_0$, given by Eq.~(\ref{C0}).

This GI approach also allows to consider less restrictive cases with constraints which are linear in the density and current.  For example, an interesting problem is: find the least expansive (in terms of the average cost) control obeying the detailed balance, i.e. with the zero current $\bm{J}=0$ for all $\tau\in[t;T]$: $\min_{\{\rho\}} C_0(\{\rho\},\{{\bm 0}\};t,{\bm x};T)$. Note that, due to the continuity equation, $\rho$ should be time-independent. The most interesting setting corresponds to the stationary force $\bm{f}$ and control $\bm{u}$ fields, when we let the system equilibrate and minimize the average cost-to-go in the long-time limit. This corresponds to minimizing with respect to $\rho$ the functional $C_{0}$, given by Eq.~(\ref{C0}), with a proper Lagrange term added to ensure normalization, in the case $\bm{J}=0$ and time-independent $\rho$ and $\bm{f}$. The above minimization leads to the following equation
\begin{eqnarray}
\label{equil-rho-var} -\frac{\kappa^{2}}{\rho}\sqrt{g}\partial_{i}\frac{h^{ij}}{\sqrt{g}}\partial_{j}\rho+ \frac{\kappa^{2}}{2\rho^{2}}h^{ij}(\partial_{i}\rho)(\partial_{j}\rho)+ U= E
\end{eqnarray}
with $E$ being the Lagrange multiplier. Here in Eq.~(\ref{equil-rho-var}) we introduced the potential
$U=\frac{1}{2}h^{ij}f_{i}f_{j}+\kappa\sqrt{g}\partial_{i}\left(h^{ij}f_{j}/\sqrt{g}\right)$.
Under the substitution, $\psi=\sqrt{\rho}$, Eq.~(\ref{equil-rho-var}), adopts a form of the Schr\"{o}dinger equation, $-(\hbar^{2}/2)\sqrt{g}\partial_{i}(h^{ij}/\sqrt{g})\partial_{j}\psi+U\psi= E\psi$,
where the newly introduced notation, $\hbar=2\kappa$, aims at making a relation of the SOC problem to the Quantum Mechanics obvious. A straightforward calculation shows that the cost-of-control rate, i.e., the average cost-to-go per unit time in the long time limit is given by $E$. Therefore, the problem of minimizing the cost of being at equilibrium is reduced to finding the ground state of a Schr\"{o}dinger equation, with the low-noise case corresponding to the quasi-classical limit.

Returning to the GI-HJB setting, note that Eq.~(\ref{HJB-u}) or Eq.~(\ref{HJB}) are nonlinear. However, in the case when the two metrics, $g$ and $h$, characterizing the statistics of noise and the cost-of-control respectively, are proportional, i.e. $g^{kj}=2\kappa q h^{kj}$, where $q$ is the scalar proportionality coefficient, the gauge-invariant HJB Eq.~(\ref{HJB}) turns into the linear equation upon substituting, $\Psi\doteq\exp(-q\Phi)$:
\begin{eqnarray}
-\partial_\tau\Psi &=&\kappa\sqrt{g}(\partial_k-qA_k)\frac{g^{kj}}{\sqrt{g}}(\partial_j-qA_j)
\Psi\nonumber\\ &+& (g^{kj}f_k(\partial_j-qA_j)\!-\!qV)\Psi.
\label{Linear}
\end{eqnarray}
The linear Eq.~(\ref{Linear}) should be solved backwards in time starting with $\Psi(T,{\bm x})=\exp(-q\varphi({\bm x}))$. This reduction from the nonlinear Eq.~(\ref{HJB}) to the linear Eq.~(\ref{Linear}) extends what has been explored in the recent papers \cite{05Kap,07Tod,08Tod,11Kap,Dj11,12Bierk,12KGO}, devoted to the so-called path-integral and Kullback-Leibler control problems, by allowing for a vector potential cost term. Note also that this transformation is akin to the so-called Cole-Hopf transformation of the Burgers equation into an auxiliary linear equation of the Schr\"{o}dinger type \cite{50Hopf}. (See also \cite{99Zar} and references therein for a discussion of the Cole-Hopf transformation applied to a HJB type of equation in the context of the dynamic programming approach to portfolio optimization in mathematical finance.)

Another interesting feature of the proportional case is that in the long-time limit the rate $C_0/(T-t)$ of the optimal average cost becomes equal, up to a simple re-scaling, to the so-called Cram\'er (or large-deviation) functional describing the negative log-probability of density and current fluctuations, $\tilde{\rho},\tilde{J}$ (defined as the expressions under the average signs on the r.h.s. of Eqs.~(\ref{rho},\ref{J}), see \cite{09CCMT} for details): $\log(-{\cal P}(\{\tilde{\rho}\},\{\tilde{\bm J}\};0,{\bm x};T))= q C_0(\{\rho\},\{{\bm J}\};0,{\bm x};T)$.
\begin{figure}
\centering
\subfigure[$f(x)=-2\cos(x)$, $\kappa=0.5$, $J=0.2$ and $\lim_{T\to\infty} {\cal C}/T\approx 2.32$]{\includegraphics[width=0.28\textwidth]{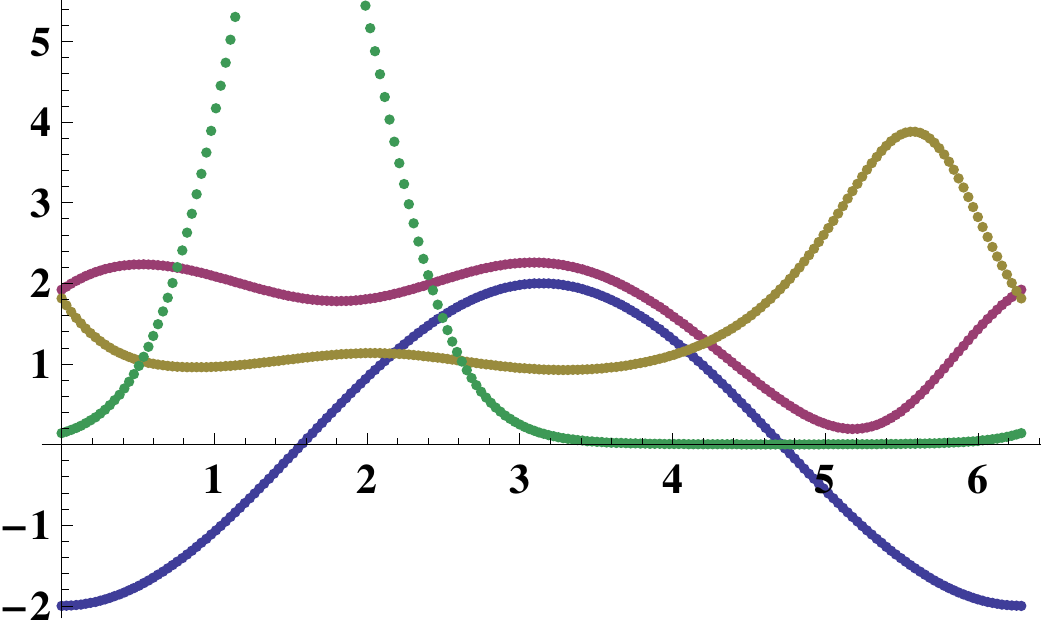}}
\subfigure[$f(x)=-2\cos(x)$, $\kappa=2.5$, $J=0.2$ and resulting in $\lim_{T\to\infty} {\cal C}/T\approx 1.81$]{\includegraphics[width=0.28\textwidth]{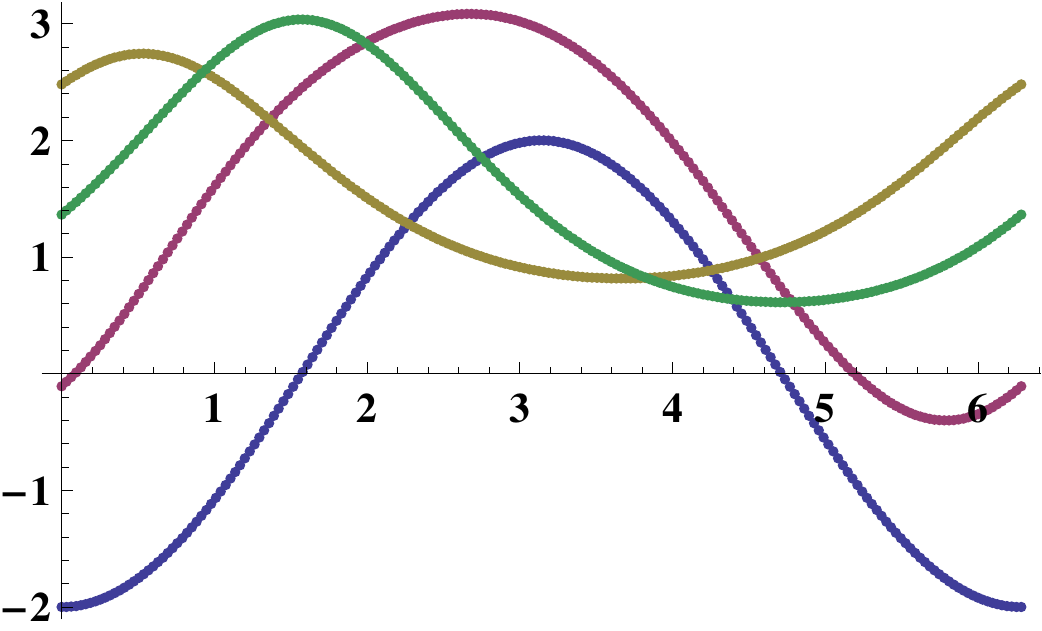}}
\subfigure[$f(x)=1-2\cos(x)$, $\kappa=0.5$, $J=0.09$ and resulting in $\lim_{T\to\infty} {\cal C}/T\approx 0.36$]{\includegraphics[width=0.28\textwidth]{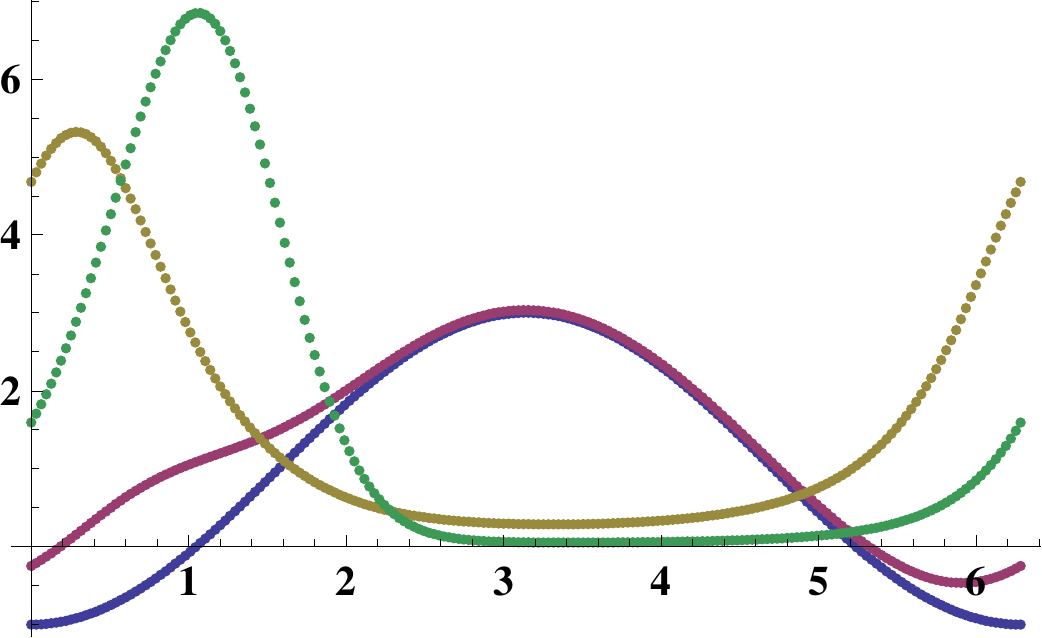}}
\caption{Three illustrative examples of the 1d (particle on the circle) ergodic control with fixed flux (zero in the cases (a,b) and nonzero in the case (c). The color coding of the curves is as follows: bright green and dark green curves show $10*\rho(x)$ in the bare (without control) and optimal control cases respectively; blue and purple curves show $f(x)$ and $f(x)+u(x)$ respectively.}
\end{figure}

An important aspect of GI-HJB equation in the context of ergodic control, i.e., the long time limit with stationary fields, is the principal capability to optimize over the fluxes. Flux over a non-contractible cycle is defined as the number of times the system goes over the cycle divided by the (long) time $T$, or equivalently as the integral over the current density $\bm{J}$ over the corresponding non-contractible $(m-1)$-dimensional surface (see \cite{09CCMT} for details). This can be done by solving the stationary version of Eq.~(\ref{HJB}) with $V=0$ and curvature free, i.e., $\partial_{i}A_{j}-\partial_{j}A_{i}=0$, vector potential, which is globally still not a gradient $A_{i}\ne \partial_{i}\varphi$.

Next we discuss our enabling ergodic case example of a particle moving along a simple circle of length $L=2\pi$ with constant $g=h=1$. Note, however, that in this special (and not fully representative) case the flux density and the current density coincide ($J(x)={\rm const}$). This case is analyzed by combining the stationary version of Eq.~(\ref{HJB-u}) with  the second expression in Eq.~(\ref{FP}), resulting in
\begin{eqnarray}
\label{1d-u-s} u^{2}/2+fu+\kappa\partial_{x}u= -E,\ -\kappa\partial_x\rho+(f+u)\rho=J,
\end{eqnarray}
where $E$ and $J$ should be treated as constants, with the periodic boundary conditions $u(x+2\pi)=u(x)$ and $\rho(x+2\pi)=\rho(x)$ imposed. It is convenient to perform analysis implicitly by fixing the value of $E$, solving Eqs.~(\ref{1d-u-s}), and thus determining the value of the flux $J$. This analysis, illustrated in Figs.~1a-c, suggests a few observations. First, the cases with and without flux-fixing control are significantly different. Since the bare (without control) flux was smaller (simply zero in the cases of Figs.~1a,b) than the resulting flux under control, the density distribution is significantly more spread out in the control case, also showing appearance of some additional structure (two local maxima in density). Second, comparing Fig.~1a and Fig.~1b, different in diffusion only, we observe that increase in diffusion spreads up the density distribution, resulting in the average decrease of the cost-to-go. We observe that the extra diffusion helps advection to boost the particle transport (flux) with less cost. Third, in the case of Fig.~1c the increasing flux leads to the control field splitting into two components, one modifying the potential (divergence free) component of the force, $f(x)$, and the other enhancing the constant/flux contribution to the force. We conclude that the optimal flux control cannot be explained simply as adjusting the gradient (potential) part of the force, or vice versa as adjusting only the constant contribution leaving the potential intact.

In this letter we focused on the physics analysis and interpretations of the density/current variational formulation of the SOC. A reader interested in related mathematically rigorous results is advised to consult with \cite{13BCCK}. The research of VYC has received support from the NSF under grant agreement no. CHE-1111350. The work at LANL was carried out under the auspices of the National Nuclear Security Administration of the U.S. Department of Energy under Contract No. DE-AC52-06NA25396. The research of JB and HJK was funded  by the FP7/2007-2013 program under the grant no. 231495.

\bibliographystyle{unsrt}
\bibliography{Bib/Control,Bib/path_int,Bib/fluid_control}

\begin{thebibliography}{10}

\bibitem{77Fle}
W.~H. Fleming.
\newblock Exit probabilities and optimal stochastic control.
\newblock {\em Applied Mathematics and Optimization}, 4(1):329--346, 1977.

\bibitem{81Mit}
Sanjoy~K. Mitter.
\newblock Non-linear filtering and stochastic mechanics.
\newblock In {\em Stochastic Systems: The Mathematics of Filtering and
  Identification and Applications}, volume~78, pages 479--503. NATO Advanced
  Study Institutes Series, 1981.

\bibitem{05Kap}
H.J. Kappen.
\newblock Path integrals and symmetry breaking for optimal control theory.
\newblock {\em Journal of Statistical Mechanics: Theory and Experiment}, page
  P11011, 2005.

\bibitem{07Tod}
E.~Todorov.
\newblock Linearly-solvable markov decision problems.
\newblock In {\em In Advances in neural information processing systems},
  volume~19, page 1369–1376. Cambridge: MIT Press, 2007.

\bibitem{08Tod}
E.~Todorov.
\newblock General duality between optimal control and estimation.
\newblock In {\em In 47th IEEE conference on decision and control}, page
  4286–4292, 2008.

\bibitem{11Kap}
H.J. Kappen.
\newblock Optimal control theory and the linear bellman equation.
\newblock {\em Inference and Learning in Dynamic Models}, pages 363--387, 2011.

\bibitem{09CCMT}
V.~Y. Chernyak, M.~Chertkov, S.V. Malinin, and R.~Teodorescu.
\newblock Non-equilibrium thermodynamics and topology of currents.
\newblock {\em Journal of Statistical Physics}, 137(1):109--147, 2009.

\bibitem{12CC}
M.~Chertkov and V.~Chernyak.
\newblock Fluid mechanics as a single-particle control.
\newblock {\em Work in Progress}, 2012.

\bibitem{Dj11}
K.~Dvijotham and E.~Todorov.
\newblock A unifying framework for linearly solvable control.
\newblock In {\em Proceedings of the 27th Annual Conference on Uncertainty in
  Artificial Intelligence (UAI-11)}, pages 179--186, Corvallis, Oregon, 2011.
  AUAI Press.

\bibitem{12Bierk}
J.~Bierkens and H.J. Kappen.
\newblock Explicit solution of relative entropy weighted control.
\newblock {\em submitted; arXiv:1205.6946}, 2012.

\bibitem{12KGO}
H.J. Kappen, V.~Gomez, and M.~Opper.
\newblock Optimal control as a graphical model inference problem.
\newblock {\em Machine Learning Journal}, pages 1--11, 2012.

\bibitem{50Hopf}
E.~Hopf.
\newblock The partial differential equation $u_t+u u_x=\mu u_{xx}$.
\newblock {\em Communications of Pure and Applied Mathematics}, 3:201--230,
  1950.

\bibitem{99Zar}
T.~Zariphopoulou.
\newblock Optimal investment and consumption models with non-linear stock
  dynamics.
\newblock {\em Math. Methods Oper. Res.}, 50(2):271--296, 1999.

\bibitem{13BCCK}
J.~{Bierkens}, V.~Y. {Chernyak}, M.~{Chertkov}, and H.~J. {Kappen}.
\newblock {Linear PDEs and eigenvalue problems corresponding to ergodic
  stochastic optimization problems on compact manifolds}.
\newblock {\em arxiv:1303.0126}, 2013.

\end{thebibliography}

\end{document}